\documentstyle[12pt]{article}
\begin{document}
\thispagestyle{empty}
\title
{\large\bf Photon absorption on a neutron-proton pair in
$^3$He\thanks{Supported by the Deutsche Forschungsgemeinschaft (SFB 201)
and the Academy of Finland}} 
\author{ {\normalsize
P. Wilhelm$^{1)}$, J.A. Niskanen$^{2)}$ and H. Arenh\"ovel$^{1)}$} \\[0.5cm]
{\normalsize\it 1) Institut f\"ur Kernphysik, Johannes
Gutenberg-Universit\"at}\\
{\normalsize\it D-55099 Mainz, Germany} \\
{\normalsize\it 2) Department of Theoretical Physics, Box 9}\\
{\normalsize\it FIN-00014 University of Helsinki, Finland }} 
\date{ }
\maketitle
\begin{abstract}
\noindent
The cross section of the reaction
$^3$He$(\gamma ,pn)p_{\mbox{\scriptsize spec}}$
in the energy range 150 -- 450 MeV has been calculated in a
quasifree two-body model and compared with a recent kinematically
complete experiment. In comparison with free deuteron disintegration
the cross section is increased by a factor of about 1.5 in reasonable
agreement with the data for photon energies above 200 MeV. At
lower energies the data deviate from the two-body behaviour and
thus suggest a more important role of the spectator requiring
a more exact treatment.
\end{abstract}
\newpage
Some recent kinematically complete photoabsorption experiments
on few-nucleon systems, notably on $^3$He and $^4$He,
 probe reaction mechanisms
involving two or three nucleons. The choice of special kinematics with all
momenta determined allows the explicit separation into two- or three-nucleon
processes and thus facilitates detailed comparisons with specific
theoretical models. Since each type of reactions can be considered separately,
the usual ambiguities are diminished.
In studies of  ``medium effects'' in certain reaction mechanisms
(albeit only few nucleons), one may consider reactions on two nucleons
in the nuclear medium as the
first frontier of interest by comparing with the free reactions on two
nucleons. Unfortunately, there is only one bound two-nucleon system to
compare with,
namely the triplet-isosinglet deuteron, while in a nuclear environment
also pairs with other
quantum numbers, notably the $^1S_0$ isotriplet, become accessible.

The TAGX collaboration has extracted the photoabsorption  cross
section of a  neutron-proton pair in $^3$He  in a
kinematically complete experiment employing tagged photons in the
intermediate energy range 150 - 450 MeV \cite{emura}. Also the angular
distributions of both  neutrons
and protons in the quasi-two-body reaction are determined for 245 MeV
photons. This
explicit extraction goes beyond other, in many respects similar,
experiments \cite{dhose,ruth}, which typically give their results in
the form of excitation functions at fixed single-nucleon angles.
In the momentum dependence of these
results it is possible to recognize a two-nucleon peak which can be compared
with theoretical predictions \cite{laget}. However, the inherent presence
of both the
significant coherent multiparticle and the single nucleon background makes
comparisons with models including just one mechanism less direct than in the
case of Ref.\ \cite{emura}.

There are two principal goals in two-nucleon absorption studies on
nuclei. On the one hand, the presence of other nucleons may open new reaction
mechanisms absent in the pure two-nucleon case. On the other hand,
it leads to a modification of the
participant two-nucleon wave function. A quasifree reaction mechanism
rests on the assumption that the explicit effects from the remaining
nucleons, i.e., spectators, can be neglected and that they retain
their initial Fermi motion distribution in the final state.
In that case information about the wave functions and correlations
in the medium can be obtained, provided the corresponding quasifree
reaction, here the two-body reaction, itself is understood.

Apart from the presence of the spectator, the main difference of a
quasideuteron in $^3$He as compared with the free deuteron is that its wave
function is more compressed to short distances with an enhanced
$D$-state probability. In the case of pion absorption, these differences cause
large effects in both the magnitude of the cross section and in spin
observables \cite{pihe}. Similar changes could be expected also in the
quasideuteron dominated part of photodisintegration. In this letter
we give a comparison  of a theoretical calculation of photoabsorption
on a bound nucleon pair with the experiment of Ref.\ \cite{emura}.
We see that the change of the wave function alone is sufficient to explain
largely the differences between the data of Ref.\ \cite{emura} and the
free deuteron photodisintegration data.

The basic ingredients of the model are identical to free deuteron
photodisintegration \cite{leide,wilhelm} and will not be repeated here,
except that the model includes the usual one-nucleon current alongwith the
Siegert operators and explicit pion and $\rho$-meson
exchange currents beyond the Siegert operators. The particularly
important contribution at intermediate energies, the M1 excitation of the
$\Delta (1232)$-isobar, is incorporated in a coupled channel
approach as developed for deuteron photodisintegration in
Refs.\ \cite{leide,wilhelm}. Allowing for a phenomenological adjustment of
the $\gamma N \Delta$-coupling in Ref. \cite{wilhelm}, this  model
gives a satisfactory
description of the energy dependence of the free cross section at
 intermediate energies. Photodisintegration of the $^1S_0 (np)$ pair
in $^3$He is coherent and
experimentally inseparable from the photon absorption on a quasideuteron.
Partly due to the lack of a strong $\Delta$ contribution in the magnetic
dipole transition in this case \cite{phohe}, it is, however, a very
minor effect in the energy
range of the experiments of Refs. \cite{emura, dhose, ruth}, but for
completeness it is included in our calculation.

The spin-isospin weights in $^3$He give for the overall probability of a
particular two-nucleon pair to form a quasideuteron $(T = 0, S = 1)$ the
value $1/2$ and $1/6$ for a $^1S_0 (np)$ pair. The remaining probability
$1/3$ is for the $^1S_0(pp)$ pair. This yields for the cross section of
absorption on an $np$ pair (the observable extracted in Ref.\ \cite{emura})
\begin{equation}
\sigma_{\mbox{\scriptsize He}}(np) = \frac{1}{2}\, \sigma_{qd} +
 \frac{1}{6}\, \sigma_s
\end{equation}
with a corresponding relation for the differential cross sections. Here
$\sigma_{\mbox{\scriptsize He}}(np),\; \sigma_{qd}$ and $\sigma_s$ are the
{\it two-body} cross sections
for absorption on $^3$He, the quasideuteron and the $^1S_0 (np)$ pair,
respectively. More details on the cross sections and also spin
observables will be given in Ref.\ \cite{phohe}.

The $S$-wave part of the quasideuteron wave function is taken from
the correlation function of Friar et al. \cite{friar}, obtained
in a Faddeev calculation using the Reid soft core potential. This
can be fitted in a smooth way by multiplying the normal deuteron
wave function by a simple radial function \cite{pihe}. This wave
function is supplemented by a $D$-state obtained by multiplying
the deuteron $D$-wave part by the same radial function giving for
the $D$-state probability $P_D (^3{\mbox{He}}) = 10.5 \%$ in line with
several Faddeev calculations.

Our main result for the energy dependence of the total cross section
is given in Fig.\ 1. The solid curves show the full results for
two-body absorption on $^3$He (the upper curve) and on the deuteron.
There is an increase by roughly a factor of $1.5$ to be compared
with the experimental observation $1.24 \pm
0.26$ \cite{emura}. However, contrary to the suggestion given in
Ref.\ \cite{emura}, the calculated enhancement is not due to the
frequently presented
argument that there are 1.5 quasideuteron pairs in $^3$He. This number
of pairs is obtained as one half (the above mentioned probability) of
the number of different ways of forming a pair among three particles.
However, for identical particles (in the isospin formalism the proton
and neutron can be considered as indistinguishable)
the total cross section must be divided by 3! \cite{gw}. Considering
a factor $1/2$ out of $1/3!$ to be included in the two-body cross
section itself, this leaves $1/3$, cancelling the ``number of pairs''
or rather the number of indistinguishable
permutations with respect to the spectator. In fact, in Fig.\ 1 the
quasideuteron result has been already reduced by the factor $1/2$
of Eq.\ (1). The pure wave function effect would have been an
enhancement by a factor of about 3 over the free reaction. Therefore,
an interpretation of the experimental result as just a
statistical factor effect in comparison with the true two-body reaction,
i.e., ignoring the wave function effect completely, would be erroneous.
The dashed curve shows the pure quasideuteron contribution leaving out
the $^1S_0 (np)$ part. The change is, however, quite small except at
lower energies,
partly reflecting the importance of the $\Delta$ contribution in
the quasideuteron process and partly being simply due to the additional
relative weight $1/3$ for the $^1S_0(np)$ component of the cross
section in Eq.\ (1).

Except for the decrease of the cross section data below 200 MeV,
which cannot be reproduced, the theoretical result is quite reasonable.
A mere scaling of the free reaction cross section also fails to
produce this behaviour at lower energies. One possible explanation is a
destructive interference with the spectator, arising at lower energies.
The distribution of the Fermi momentum could allow for a significant
contribution from this interference up to 200 MeV/c nucleon  momenta.
Furthermore,
the calculation slightly overestimates the cross section.
This overestimation above 250 MeV could be compensated for by a slight
downshift in energy of the theory, which could be interpreted as a
stronger $\Delta$ attraction in the three-nucleon system than in the
two-body system.  However, the calculation is still nearly within
the experimental errors and, in fact, Ref.\ \cite{dhose}
gives a somewhat larger cross section for two-nucleon absorption
 with $\sigma_{\mbox{\scriptsize He}} / \sigma_d = 1.68 \pm
0.07$ obtained from the proton excitation function in $(\gamma, p)$
reactions.

In Fig.\ 2 we present a comparison of the angular distribution
at 245 MeV between our model results and the data of Ref.\ \cite{emura}
for neutrons (full curve, full circles) and protons (dotted curve,
open circles). The comparison of the neutron results is more direct,
since it avoids a possible participation of the spectator. Considering
the significant error limits in the data, the model agrees with the
experiment quite satisfactorily. The proton distribution is
easier to measure and has smaller errors. However, the experimental
indistinguishability of the spectator would, in principle, make its more
explicit treatment necessary also in the model cross section.
In the dotted curve we have simply added incoherently the two-body
proton cross section $d \sigma /d \Omega_p$ and an isotropic
spectator cross section $\sigma_{\mbox{\scriptsize He}}(np)/4\pi$.
The model does, indeed, give a more forward peaked proton distribution
than the neutron one in agreement with the data. However, the proton cross
section is too high, which is also reflected in the total cross section.

In summary, we have calculated quasideuteron photodisintegration with a
modified deuteron wave function and compared to the quasifree
photoabsorption on an $np$ pair measured in Ref.\ \cite{emura}.
The embedding of the quasideuteron in $^3$He leads to a probability
factor $1/2$ in the cross section. In addition, we have included
the experimentally indistinguishable absorption on the $^1S_0 (np)$ pair
 in $^3$He. Although there appears some slight
deviation in the proton cross section, where also the spectator may
contribute, the overall results are satisfactory and largely support
the quasifree assumption for the proper kinematic conditions. The
low energy behaviour requires more detailed work.  Purely compressing
the two-body wave function to shorter distances in $^3$He causes an
increase in the cross section
by a factor of about 3. This is reduced to about $1.5$ by the above
discussed probability factor $1/2$ for finding a quasideuteron in $^3$He.
One can, indeed, expect an increase in the transition matrix elements
for both one- and two-body contributions to the electromagnetic current.
The latter are shorter-ranged and their contributions
obviously benefit from the initial pairs concentrated at shorter distances.
The nucleon scattering function in turn makes the one-body current
amplitude
essentially a Fourier transform of the bound pair wave function. At the
rather high momentum transfers relevant here also this becomes larger.
This interplay of different mechanisms versus properties of the
bound pair wave function
will be discussed in more detail in a forthcoming paper \cite{phohe}.
\\[1.cm]

\centerline{\small\bf ACKNOWLEDGEMENTS}

J.A.N. thanks the Institut f\"ur Kerphysik of the University of Mainz
for hospitality.
P.W. acknowledges the hospitality of the Resarch Institute for
Theoretical Physics of the University of Helsinki.

\newpage
\centerline{\bf Figure captions}

\noindent{\bf Fig. 1.} The total cross section of the quasifree
reaction  $^3{\mbox{He}}(\gamma ,pn)p_{\mbox{\scriptsize spec}}$. The solid
curves present the full result for $^3$He (upper) and deuteron (lower).
The dashed curve describes photon absorption on the pure quasideuteron in
$^3$He without the singlet contribution and the dotted absorption
on the $^1S_0$ pair alone. The data are from Ref.\ \cite{emura}.\\

\noindent{\bf Fig. 2.} The differential cross section of two-nucleon
photodisintegration in $^3$He at 245 MeV photon energy. The solid
curve and full circles correspond to the neutron, the dotted curve
and open circles to the proton cross section.
The data are from Ref.\ \cite{emura}.
\end{document}